\documentclass[runningheads]{llncs}
\usepackage{graphicx}
\usepackage{amssymb}
\usepackage{tikz}
\usetikzlibrary{quantikz2}
\usepackage{caption}
\usepackage{subcaption}
\usepackage{nicefrac}
\usepackage{hyperref}
\usepackage{blindtext}
\begin{document}
\title{Sequential Hamiltonian Assembly: Enhancing the training of combinatorial optimization problems on quantum computers \thanks{This is an extended version of our previously published work \cite{Stein_Roshani}, additionally providing a new partitioning strategy, a new merger between QAOA and SHA, and an evaluation of SHA in the Max-Cut optimization problem.}}
\titlerunning{Sequential Hamiltonian Assembly}
%
\author{Navid Roshani\inst{1} \and 
Jonas Stein\inst{1} \and
Maximilian Zorn\inst{1} \and
Michael Kölle\inst{1} \and \\
Philipp Altmann\inst{1} \and
Claudia Linnhoff-Popien\inst{1}}
\authorrunning{N. Roshani et al.}
%
\institute{LMU Munich, Germany \\
\email{n.roshani@icloud.com} \\
\email{jonas.stein@ifi.lmu.de}}
\maketitle              
\begin{abstract}
A central challenge in quantum machine learning is the design and training of parameterized quantum circuits (PQCs). Much like in deep learning, vanishing gradients pose significant obstacles to the trainability of PQCs, arising from various sources. One such source is the presence of non-local loss functions, which require the measurement of a large subset of qubits involved. To address this issue and facilitate parameter training for quantum applications using global loss functions, we propose Sequential Hamiltonian Assembly (SHA). SHA iteratively approximates the loss by assembling it from local components. To further demonstrate the feasibility of our approach, we extend our previous case study by introducing a new partitioning strategy, a new merger between QAOA and SHA, and an evaluation of SHA onto the Max-Cut optimization problem. Simulation results show that SHA outperforms conventional parameter training by 43.89\% and the empirical state-of-the-art, Layer-VQE by 29.08\% in the mean accuracy for Max-Cut. This paves the way for locality-aware learning techniques, mitigating vanishing gradients for a large class of practically relevant problems.

\end{abstract}

\section{\uppercase{Introduction}}\label{sec:introduction}
One of the most promising avenues towards achieving early quantum advantage lies in the realm of quantum machine learning, specifically through the use of parameterized quantum circuits (PQCs) \cite{Cerezo2021}. PQCs are widely regarded as the quantum counterpart to artificial neural networks, functioning as arbitrary function approximators equipped with trainable parameters \cite{PhysRevA.103.032430}. From a mathematical perspective, PQCs are parameterized linear functions that operate within an exponentially high-dimensional Hilbert space, proportional to the number of qubits employed. The capacity to efficiently perform certain computations within this large space provides a basis for provable quantum advantage \cite{10.1145/237814.237866} \cite{deutsch1992rapid}.\\
A fundamental distinction in the gradient-based training processes between classical artificial neural networks (ANNs) and PQCs lies in their efficiency. In classical ANNs, gradient calculations are invariant with respect to the number of parameters; however, in PQCs, the runtime complexity of gradient calculations scales linearly with the number of parameters \cite{PhysRevA.98.032309}. Given that a gradient is simply the expectation value derived from the probabilistic measurement of a quantum circuit, its error-dependent runtime scaling is $\mathcal{O}\left(1/\epsilon\right)$, in contrast to the classical scaling of $\mathcal{O}\left(\log\left(1/\epsilon\right)\right)$ \cite{knill2007optimal}. This inefficiency becomes particularly pronounced in scenarios involving vanishing gradients, a common issue in PQCs \cite{McClean2018}. Notably, gradients in PQCs can diminish exponentially with the increasing number of qubits \cite{McClean2018}, unlike the scenario in classical neural networks where gradient vanishing is typically associated with the number of layers \cite{bradley2010learning,pmlr-v9-glorot10a}. \\
Consequently, much research has focused on exploring the causes of vanishing gradients in PQCs, identifying four possible factors:
\begin{enumerate}
    \item \textbf{Expressiveness}: Larger reachable subspaces of the Hilbert space increase the likelihood of vanishing gradients~\cite{PRXQuantum.3.010313}.
    \item \textbf{Locality} of the measurement operator associated with the loss function: The more qubits that need to be measured, the higher the likelihood of vanishing gradients \cite{cerezo2021cost,uvarov2021barren,kashif2023impact}.
    \item \textbf{Entanglement} in the input: More entangled or random initial states increase the likelihood of vanishing gradients \cite{McClean2018,cerezo2021cost}.
    \item Hardware \textbf{noise}: Greater noise and more varied noise types in hardware increase the likelihood of vanishing gradients \cite{wang2021noise,stilck2021limitations}.
\end{enumerate}

In recent developments, mathematical frameworks have been proposed that unify the theoretical underlying causes of vanishing gradients in PQCs. These frameworks facilitate the quantification of the extent to which vanishing gradients are present in a given PQC \cite{ragone2023unified} \cite{fontana2023adjoint}. Drawing parallels to the vanishing gradient problem in classical machine learning, various techniques are under active investigation to mitigate this issue in PQCs (see \cite{ragone2023unified} for an overview). Among the explored strategies are advanced methods for parameter initialization \cite{NEURIPS2022_7611a3cb} and the application of adaptive, problem-specific learning rates \cite{PRXQuantum.3.020365}.\\
In our previous work \cite{Stein_Roshani}, we introduced an innovative approach termed \emph{Sequential Hamiltonian Assembly} (SHA) to enhance the efficiency of parameter training in PQCs, specifically addressing the challenge of locality. SHA leverages the structure of commonly employed measurement operators $\hat{H}$: $\hat{H}=\sum_i \hat{H}_i$, where $\hat{H}_i$ is a \emph{local} Hamiltonian. Here, locality means that the operator acts non-trivially only on a small subset of qubits, thereby influencing the loss function primarily through the outcomes of these few qubits. A significant class of problems that exhibit this property and are promising candidates for demonstrating quantum advantage are combinatorial optimization problems \cite{10.3389/fphy.2014.00005} \cite{PhysRevX.8.031016} \cite{pirnay2023inprinciple}.\\
Inspired by iterative learning techniques such as layerwise learning in (quantum) machine learning and iterative rounding in optimization, SHA initiates the process with a partial sum of the measurement operator $\hat{H}$ (for example, starting with just $\hat{H}_1$) and progressively incorporates additional terms until the complete measurement operator $\hat{H}$ is fully \emph{assembled}. This iterative approach to approximating the loss function allows the training to commence with a local measurement operator, thereby facilitating the discovery of suitable initial parameters outside of potential barren plateaus, which are regions where the gradient diminishes significantly. As the process advances, this method aims to continuously circumvent barren plateaus by incrementally integrating more terms until the entire, often global, measurement operator is employed.\\
As a proof of principle, we conducted in our previous work a case study for the problem of graph coloring using state-of-the-art PQC-based approaches: the Variational Quantum Eigensolver (VQE) and the Quantum Approximate Optimization Algorithm (QAOA). We chose the problem of graph coloring because it has a complex loss function that challenges standard parameter training approaches and allows for a comparison of different assembly approaches. Our evaluation showed a significant improvement in solution quality when using SHA compared to standard gradient descent-based training and comparable state-of-the-art approaches from related work. We now extent this case study by introducing several new contribution. We introduce a new clustering-based SHA strategy, which will be compared to the previous ones. Furthermore, our previous results showed interesting findings for the solution accuracy of the most likely shot for the QAOA. Toimprove upon this quantitiy and the overall accuracy in general, we investigate the combination of SHA with the QAOA. Additionally, to further validate the advantages of SHA, it is applied onto the well-known Max-Cut problem.\\
The remainder of this paper is structured as follows. Section \ref{sec:preliminaries} provides necessary theory for parameter training in PQCs, as well as the intuitions underlying the VQE and the QAOA. Afterwards, section \ref{sec:relatedwork} discusses two different approaches from related work, that can be used to accelerate parameter training. In section \ref{sec:SHA}, we present the theory behind Sequential Hamiltonian Assembly (SHA). The experimental setup used for evaluation is then described in section \ref{sec:methodology}, before evaluating the results in section \ref{sec:results}. Finally, we conclude our findings and propose possible future work in section \ref{sec:conclusion}.

\section{\uppercase{Preliminaries}}\label{sec:preliminaries}
In this section, we present the fundamental theory underpinning the parameter training in PQCs. Additionally, we introduce the algorithms employed for their evaluation.

\subsection{Training parameterized quantum circuits}
\label{subsec:paramtraining}
Similar to classical machine learning, most practical parameter training techniques for PQCs use gradient-based methods. The key to calculating the gradients of a PQC \( U (\theta, x) \)—where \( U \) is a unitary matrix acting on all \( n \) qubits, \( x \in \mathbb{C}^k \) represents the data input, and \( \theta \in \mathbb{R}^m \) denotes the parameters—is the \textit{Parameter Shift Rule}. This rule leverages the periodic nature of single-qubit gates, exploiting the fact that all PQCs can be decomposed into parameterized single-qubit gates and non-parameterized two-qubit gates \cite{nielsen_chuang_2010}. Similar to how \( \frac{d}{dx} \sin(x) = \sin(x + \pi/2) \), it can be shown that for all \( i \):
\[
\partial_{\theta_i} U (\theta,x)|\psi\rangle = \frac{U (\theta_+,x)|\psi\rangle - U (\theta_-,x)|\psi\rangle}{2},
\]
where $\theta_{\pm} := (\theta_1, \ldots, \theta_{i-1}, \theta_i \pm \pi/2, \theta_{i+1}, \ldots, \theta_m)$ and \(|\psi\rangle\) being an arbitrary initial state \cite{PhysRevA.98.032309} \cite{PhysRevA.99.032331}. This approach enables the gradient calculation to be as efficient as the forward pass for each parameter. Additionally, by parallelizing the process with multiple QPUs, it achieves the same runtime complexity as the backward pass in classical ANNs, assuming errors are neglected.

\subsection{VQE}

The Variational Quantum Eigensolver (VQE) is a quantum optimization algorithm designed to approximate the ground state of a given Hamiltonian \( \hat{H} \), which is the eigenvector corresponding to its lowest eigenvalue \cite{Peruzzo2014}. The VQE utilizes a PQC \( U (\theta) \) and is rooted in the variational method. This method involves iteratively adjusting a function (in this case, \( f : \theta \mapsto U (\theta) |0\rangle \)) to approximate the minimum of another function (here, \( g : |\phi\rangle \mapsto \langle \phi | \hat{H} | \phi \rangle \))\cite{lanczos2012variational}.\\
While the VQE was initially developed for chemical simulations, it can also be applied to a wide range of combinatorial optimization problems by following these steps:

\begin{itemize}

\item[1.] Encode the problem: Represent the domain of the combinatorial optimization function \( h : X \to \mathbb{R} \) in binary format, defining a mapping \( e : X \to \{0,1\}^n \). This enables the formulation \( h(x) = \langle e(x) | \hat{H} | e(x) \rangle \), where \( \hat{H} \) is a diagonal matrix with eigenvalues \( h(x) \) associated with eigenvectors \( |e(x)\rangle \). Thus, finding the ground state of \( \hat{H} \) translates to finding the global minimum of \( h \).

\item[2.] Select quantum circuit: Choose a quantum circuit architecture that defines the function approximator \( U (\theta) \).

\item[3.] Initialize state and parameters: Start with an initial state \( |\psi\rangle \) (commonly \( |0\rangle \) to simplify state preparation) and initial parameters \( \theta \) (e.g., \( \theta_i = 0 \) for all \( i \)).

\item[4.] Optimize parameters: Choose an optimizer for training the parameters.
\end{itemize}
Despite extensive research \cite{Du2022} \cite{sim2019expressibility}, identifying a suitable and efficient circuit architecture remains the most challenging part of implementing the VQE in practice.

\subsection{The Quantum Approximate Optimization Algorithm}

The Quantum Approximate Optimization Algorithm (QAOA) is a variant of the VQE that integrates principles from Adiabatic Quantum Computing (AQC) to create effective quantum circuits \cite{farhi2014quantum}. AQC is a computational framework equivalent to the standard quantum gate model \cite{10.1109/FOCS.2004.8} and is grounded in the adiabatic theorem. This theorem asserts that a quantum system will remain in its ground state if changes to the Hamiltonian occur slowly enough \cite{Born1928}. By linking ground state problems with combinatorial optimization, AQC provides a robust approach to solving optimization problems.\\
In AQC, the computation is governed by a time-dependent Hamiltonian \( \hat{H}(t) = (1-t)\hat{H}_M + t\hat{H}_C \), where \( t \) progresses from 0 to 1. Here, \( \hat{H}_M \) is the Hamiltonian whose ground state represents the initial state, and \( \hat{H}_C \) is designed to correspond to the optimization problem. The straightforward preparation of an initial state for \( \hat{H}_M \) (e.g., \( |+\rangle^{\otimes n} \) with \( \hat{H}_M = -\sum_{i=1}^n \sigma_i^x \)) and the fact that \( \hat{H}_C \) can be decomposed into a sum of polynomially many local Hamiltonians make AQC a powerful tool for many optimization problems.\\
The QAOA adapts this continuous-time evolution of AQC into a discrete framework suitable for quantum computers, which use quantum gates for computation. According to the adiabatic theorem, the speed of evolution is constrained by the gap between the smallest and second-smallest eigenvalues of \( \hat{H}(t) \). To exploit faster evolution, QAOA introduces parameters to control the evolution speed. The quantum circuit in QAOA is:
\[
U(\beta, \gamma) = U_M(\beta_p) \cdot U_C(\gamma_p) \cdot \ldots \cdot U_M(\beta_1) \cdot U_C(\gamma_1) \cdot H^{\otimes n}
\]
where \( U_M(\beta_i) = e^{-i \beta_i \hat{H}_M} \), \( U_C(\gamma_i) = e^{-i \gamma_i \hat{H}_C} \), and \( p \) is a positive integer. As \( p \) approaches infinity, \( U(\beta, \gamma) \) increasingly resembles the AQC model. The parameters \( \beta_i \) and \( \gamma_i \) are set as \( \beta_i = 1 - i/p \) and \( \gamma_i = i/p \), respectively.\\
Although QAOA (and its various adaptations) often provides leading results compared to other quantum optimization techniques \cite{blekos2023review}, its performance is highly sensitive to the number of local Hamiltonians \( \hat{H}_i \) in \( \hat{H}_C = \sum_i \hat{H}_i \) and their compatibility with the hardware topology. The need for multiple applications of \( U_C \) restricts the number of feasible discretization steps \( p \), thereby affecting solution quality, which scales with \( p \). As a result, other VQE-based PQCs often surpass the QAOA on near-term quantum computers, despite the QAOA's theoretical potential to find optimal solutions given enough time \cite{9669165} \cite{Skolik2021}.
\section{\uppercase{Related Work}}\label{sec:relatedwork}
In our previous work \cite{Stein_Roshani}, we introduced Sequential Hamiltonian Assembly and showed its advantages compared to other Quantum Learning methods. To compare our approach with other methods for enhancing parameter training in VQE-based PQCs, we now introduce two prominent techniques  shortly: Layerwise learning and Layer-VQE. To the best of our knowledge, no other baselines have been proposed that are more similar to our methodology in terms of iteratively guiding the parameter learning process while aiming to avoid barren plateaus.

\subsection{Layerwise Learning}
\label{subsec:ll}
Inspired by classical layerwise pretraining strategies in deep learning (see \cite{NIPS2006_5da713a6}), \cite{Skolik2021} demonstrated that iteratively training a subset of parameters in PQCs can significantly enhance solution quality. Their approach assumes a layered structure for the PQC, which is common in most literature, and involves two phases of training.\\
In the first phase, parameters are trained while sequentially assembling the PQC layer by layer:

\begin{itemize}
    \item[1.] Start with a PQC consisting of the first $s$ layers, initializing all parameters to zero to avoid barren plateaus.
    \item[2.] Train the parameters for a predefined number of optimization steps.
    \item[3.] Add the next $p$ layers, fixing the parameters of all but the last added q layers, and initialize all new parameters to zero.
    \item[4.] Train the parameters of the last added $q$ layers for a predefined number of optimization steps.
\end{itemize}
This process continues until adding new layers no longer improves the solution quality or a specified maximum depth is reached. The variables $s$, $p$, and $q$ are hyperparameters that may need to be adjusted based on the specific problem requirements. \\
In the second phase, another round of parameter training is conducted with the fully assembled circuit. A fixed fraction of layers, $r$, is trained in a sliding window manner while the rest of the circuit's parameters remain fixed. Each subset of layers is trained for a fixed number of optimization steps. \\
By keeping the number of optimization steps low for each part of the training, overfitting is prevented, and the overall training duration is kept manageable. Subsequent studies, such as \cite{PhysRevA.103.032607}, have shown a lower bound on the subset size of simultaneously trained layers for effective training. Nonetheless, layerwise learning has shown significantly lower generalization errors in relevant applications like image classification \cite{Skolik2021}.

\subsection{Layer VQE}

Building on the insights from \cite{Skolik2021} and \cite{PhysRevA.103.032607} (see section \ref{subsec:ll}), \cite{9669165} proposed the iterative parameter training approach Layer-VQE, which essentially mirrors a special case of layerwise learning. The core idea of Layer-VQE is that each layer must equal an identity operation when its parameters are set to zero. This ensures that adding a new layer does not change the circuit's output state, allowing the search in the solution space to continue from the previously optimized solution. Unlike typical layerwise learning, Layer-VQE omits the second phase (i.e., $r = 0$) and trains all previously added layers simultaneously by choosing $q$ to cover all layers. To limit the number of new parameters added at each step, only one layer is added in each iteration (i.e., $p = 1$). Additionally, Layer-VQE includes an initial layer of parameterized $R_y$ rotations on every qubit.\\
According to the large-scale evaluation by \cite{9669165}, Layer-VQE can outperform QAOA in terms of solution quality and circuit depth for specific optimization problems, particularly on noisy hardware.
\section{\uppercase{Sequential Hamiltonian Assembly}}\label{sec:SHA}
We now go over our method introduced in our previous work \cite{Stein_Roshani}: The Sequential Hamiltonian Assembly (SHA) approach, aimed at improving parameter training of PQCs on global cost functions. Inspired by methods like layerwise learning, where the quantum circuit is assembled layer by layer, we propose assembling the in general global Hamiltonian \( \hat{H} = \sum_{i=1}^N \hat{H}_i \) by iteratively combining its predominantly local components \( \hat{H}_i \). This strategy mirrors successful combinatorial optimization methods that start with a simplified version of the cost function and gradually reassemble the original cost function by removing relaxations, see \cite{bansal:LIPIcs:2014:4827}.\\
Below are the steps outlining the SHA concept for a given PQC and decomposable Hamiltonian \( \hat{H} = \sum_{i=1}^N \hat{H}_i \):

\begin{itemize}
    \item[1.]Determine the order in which Hamiltonians will be added in each iteration, creating a partition \( P = \{P_1, \ldots, P_M\} \) where \( P_i \subseteq \{1, \ldots, N\} \) and \( \bigcup_{j=1}^M P_j = \{1, \ldots, N\} \).
    \item[2.] Set a maximum number of parameter optimization steps per iteration \( s_j \), ensuring it is low enough to avoid overfitting.
    \item[3.] Optimize the parameters of the PQC wrt. the Hamiltonian \( \sum_{i \in \bigcup_{j=1}^k P_j} \hat{H}_i \) for each \( k \in \{1, \ldots, N\} \) iteratively for a maximum of \( s_k \) steps. 
\end{itemize}
As demonstrated in our evaluation, the assembly strategy significantly impacts solution quality. We evaluate four different approaches:

\begin{enumerate}

    \item \textbf{Random} Use equally sized, non-overlapping partitions, assigning each \( \hat{H}_i \) randomly.
    \item \textbf{Sequential } Use equally sized, non-overlapping partitions, assigning Hamiltonians in the order provided.
    \item \textbf{Clustering} For graph based applications, cluster the underlying graph to assign Hamiltonians to the corresponding cluster.
    \item \textbf{Problem inspired} Use a partitioning strategy where terms in each partition share a common, problem-specific property.
\end{enumerate}
Practically, given Hamiltonians can be divided into numerous sub-Hamiltonians \( \sum_{i=1}^N \hat{H}_i \), progressing with one \( \hat{H}_i \). However this approach might be too slow for practical use. For our purposes, and due to computational constraints, \( M \leq 10 \) showed decent results. These four approaches offer varying degrees of problem-specific information: \emph{Random} provides no information, \emph{Chronological} often provides some, \emph{Cluster} relies on the available information embedded in the graph and \emph{Problem inspired} uses all available knowledge to solve the problem iteratively, as demonstrated in the following examples of graph coloring and Max-Cut.\\
Graph coloring is a satisfiability problem where each node is assigned a color such that no adjacent nodes share the same color. While optimization seeks to minimize the number of colors, we focus on the satisfiability version due to its complex structural properties and ease of evaluation. To conserve computational resources, we use the Hamiltonian formulation from \cite{9259934}, which requires the least qubits:

\[ \sum_{(v,w) \in E} \sum_{a \in \mathbb{B}^m} \prod_{l=1}^m \left( 1 + (-1)^a \sigma_{v,l}^z \right) \left( 1 + (-1)^a \sigma_{w,l}^z \right) \]

This formulation is valid for problems where the number of colors \( k \) is a power of two (i.e., \( \exists m \in \mathbb{N}: 2^m = k \)), ensuring minimal computational requirements. This Hamiltonian can be decomposed into at most \( |E| \cdot k \) Pauli terms with at most \( |E| \cdot \left\lceil \log_2 k \right\rceil \)-local terms. We propose a node-wise approach, creating \( |V| \) partitions \( P_j \) that contain all Pauli term indices involving the node \( v_j \in V \). \\
Max-Cut on the other hand is a classical combinatorial optimization problem that aims to partition the vertices of a graph into two disjoint subsets such that the number of edges between the subsets is maximized. Its Hamiltonian is given by:
$$ \sum_{(v,w) \in E} (1-\sigma^z_v \sigma^z_w)$$
Since the Max Cut problem is also graph-based, the node-wise strategy will be applied to it as well. \\
For both problems our results show that increased problem instance information significantly improves solution quality. Thus, SHA exemplifies how to address the challenge of training with global cost functions by iteratively assembling them from local subproblems in an informed manner.
\section{\uppercase{Experimental Setup}}\label{sec:methodology}

In this section, we explain and justify our selection of problem instances, PQC architectures, and hyperparameters used in the subsequent evaluation.

\subsection{Generating Problem Instances}

To generate unbiased, statistically relevant problem instances, we employ the Erdős-Rényi-Gilbert model for random graphs \cite{10.1214/aoms/1177706098}. This model enables us to create graphs with a fixed number of nodes while varying the number of edges, thereby producing graphs of different difficulty levels. Generally, the difficulty of solving the graph coloring problem for a fixed number of colors in a random graph increases with the number of edges \cite{PhysRevE.76.031131}. To quantify this difficulty, we use the percentage of correct solutions in the search space, a common metric for satisfiability problems.

Table \ref{tab:graphs} presents the dataset generated using this approach. Here, \( p \) represents the probability of any two nodes being connected by an edge, \( r \) denotes the percentage of correct solutions in the search space, \( s \) is the absolute number of correct solutions and \textit{E} is the optimal Energy for Max-Cut. To balance computational effort and graph size, all instances for both problems involve 8 nodes and 4 colors, requiring \( 8 \cdot \log_2(4) = 16 \) qubits and 8 qubits respectively, and yielding a search space size of \( 4^8 = 65535 \). As our evaluation shows, these graphs vary in difficulty based on \( p \) and \( s \), with graph 9 being the hardest and graph 5 the easiest.

\begin{table}
    \centering
    \begin{tabular}{c||c|c|c|c|c}
        Graph ID & \( p \) & \( r \) & \( s \) & \( E \) & Seed \\ \hline \hline
        1  & \ 0.30 \ & \  1.025\% \ & \ 672 \  & \  12 \  & \ 7 \ \\
        2  & 0.55 & 1.501\% & 984 & 11 & 8 \\
        3  & 0.40 & 1.025\% & 672 & 11 & 9 \\
        4  & 0.40 & 1.428\% & 936 & 9 & 10 \\
        5  & 0.35 & 3.369\% & 2208 & 9 & 11 \\
        6  & 0.30 & 2.051\% & 1344 & 10 & 12 \\
        7  & 0.35 & 3.223\% & 2112 & 10 & 13 \\
        8  & 0.50 & 0.879\% & 576 & 10 & 14 \\
        9  & 0.90 & 0.037\% & 24 & 15 & 15 \\
        10 & 0.40 & 0.659\% & 432 &  12 & 16 \\
    \end{tabular}
    \caption{Generated graph problem instances using the \texttt{fast\_gnp\_random\_graph} function from \texttt{networkx} \cite{hagberg2008exploring}. Each graph was verified to be fully connected.}
    \label{tab:graphs}
\end{table}

\subsection{Selecting Suitable Circuit Layers}

To evaluate a variety of PQC architectures, we reference the extensive list provided in \cite{https://doi.org/10.1002/qute.201900070}, which includes diverse PQC structures. We extend these architectures to the required 16 qubits by identifying and expanding the underlying design principles (e.g., ladder, ring, or triangular entanglement layers, as well as single-qubit rotation layers). Given computational constraints, we conducted a preliminary study to select a subset of architectures based on the following criteria: (1) significantly better-than-random performance, (2) limited number of parameters (to reduce training time), and (3) architectural variance. Consequently, we chose circuits 1, 3, 8, 12, 13, 16, and 18 for our evaluation, having excluded others based on these criteria. The preliminary study indicated that these circuits offer comparable solution quality, with circuit 12 performing slightly worse, possibly due to its layers' non-identity property when all parameters are zeroed.

\subsection{Hyperparameters}

As outlined in Sections \ref{sec:preliminaries}, \ref{sec:relatedwork}, and \ref{sec:SHA}, both the baseline methods and SHA have crucial hyperparameters that we now specify. Most approaches require layerwise structured PQCs, prompting a preliminary study to determine the necessary number of circuit layers for solving the problem instances. We found that using three layers is optimal, as additional layers did not significantly improve solution quality. For layerwise learning, we set \( s = 1 \), \( p = 1 \), and \( q = 1 \), to train only one layer at a time initially, and \( r = 1 \) to train the full PQC in the second phase, as per the approach in \cite{PhysRevA.103.032607}.\\
For the Cluster strategy we employ the KMeans algorithm \cite{IKOTUN2023178} with preset hyperparameters implemented in \texttt{sklearn} \cite{scikit-learn}. \\
Furthermore, for all methods, we employed the COBYLA optimizer \cite{COBYLA1}, due to its proven efficiency for similarly sized problems \cite{Joshi_2021} \cite{9345015}. We set the maximum number of optimization steps to 4000 for all runs, with a minimal progress requirement of 0.8 per step during partial parameter training, and \( 10^{-6} \) during full parameter training to prevent overfitting. Additionally, we used a shot-based circuit simulator with 200 shots for each circuit execution, ensuring accurate results that reflect real quantum hardware conditions.
\section{\uppercase{Results}}\label{sec:results}
To evaluate our approach, we first focus on the Graph coloring probplem. We compare the proposed assembly strategies and then review the solution quality of SHA. We also demonstrate how SHA can be effectively combined with other quantum learning methods, such as layerwise learning and Layer-VQE. Afterwards, we compare the time complexity of all discussed approaches. Finally, we go over to the Max-Cut problem and compare SHA to the other methods. To ensure statistical significance, all experiments are averaged over five seeds.

\subsection{Comparing Assembly Strategies}
\begin{figure}[ht]
    \centering
    \includegraphics[width=\columnwidth]{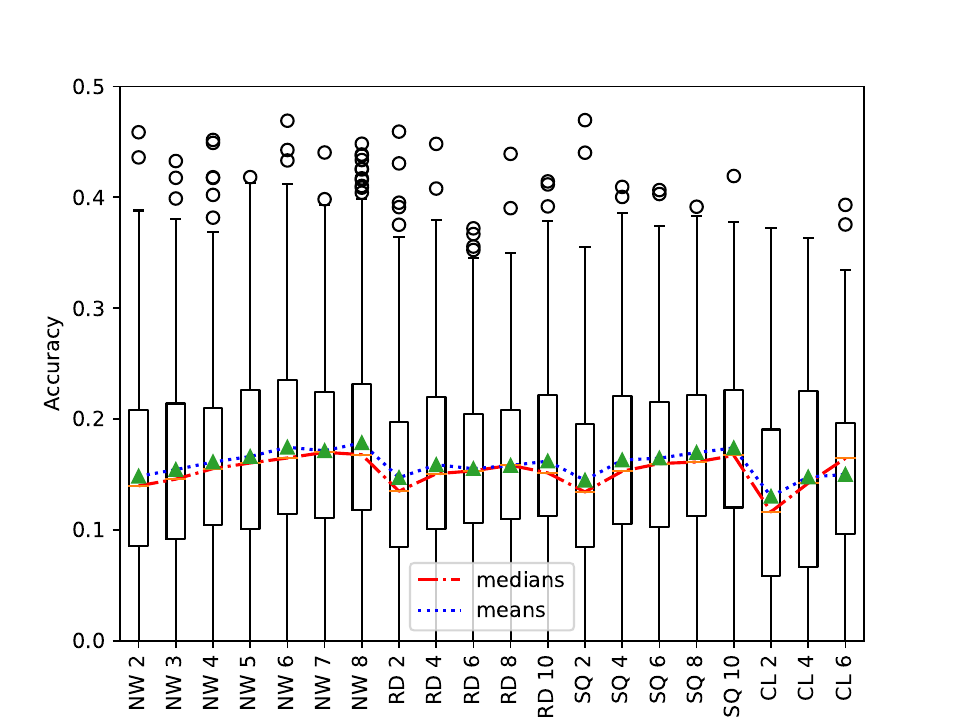}
    \caption{Accuracy over all graphs and circuit architectures per assembly strategy.}
    \label{fig:assembly-strategies}
\end{figure}

As discussed in Section~\ref{sec:SHA}, SHA's performance depends on the Hamiltonian assembly strategy. We evaluate three strategies: (1) random, (2) chronological, and (3) problem-inspired. The results in Figure~\ref{fig:assembly-strategies} indicate that the cluster strategy (CL $i$) performed the worst when partitioning into $i \in \{2,4,6\}$ partitions parts. Afterwards, the random strategy (RD $i$) follows with slightly better results to begin with. The chronological approach (SQ $i$) performed better and approached the performance of the problem-informed, nodewise strategy (NW $j$), where $j \in \{2,\ldots,8\}$ denotes the number of connected subgraphs used in the assembly. Clearly, a problem-informed strategy yields better results from the present results. The chronological approach also shows an explicit progression, where more structured partitions lead to better outcomes. For the cluster strategy the graphs itself didn't yield enough usable information, due to the comparably simplicity of the graphs itself. We suggest future work to explore higher values for $i$, expecting further improvements, especially for larger problem instances. Furthermore, clustering on larger graphs should be investigated further once clearly distinguishable clusters are present. To ensure runtime comparability with the baselines, we use the nodewise approach for the remainder of this evaluation.

\subsection{Solution Quality}
\begin{figure*}[ht]
    \begin{center}
    \begin{subfigure}[t]{0.49\textwidth}
         \centering
         \includegraphics[width=\columnwidth]{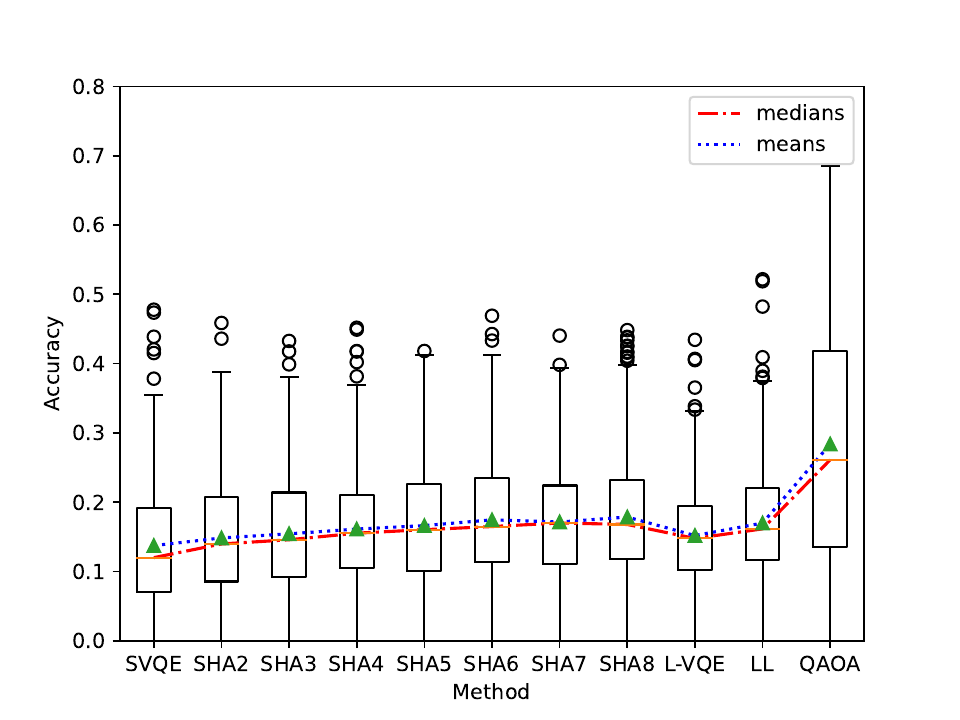}
         \caption{Ratio of correct solutions.}
         \label{fig:accuracy}
     \end{subfigure}
     \begin{subfigure}[t]{0.49\textwidth}
         \centering
         \includegraphics[width=\columnwidth]{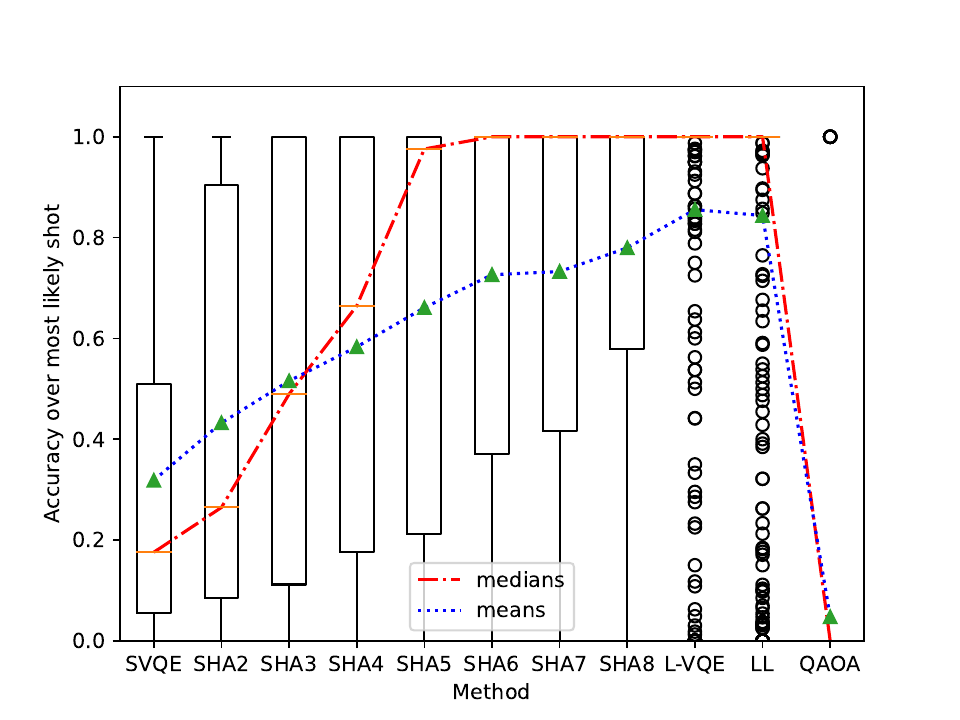}
         \caption{Ratio of correct solutions of the most likely shot, averaged over the last 2\% of optimization steps.}
         \label{fig:accuracy-most-likely-averaged}
     \end{subfigure}
    \caption{Solution quality averaged over all graphs and circuit architectures for all approaches. \cite{Stein_Roshani}}
    \label{fig:solution-quality}
    \end{center}
\end{figure*}

To evaluate solution quality, we examine two properties: overall accuracy and accuracy of the most likely solution. While most literature focuses on overall accuracy, investigating the most likely solution reveals the focal point of the identified solution set. Interestingly, our experiments show that these two properties do not necessarily coincide. High accuracy in one does not imply high accuracy in the other, as shown in Figure~\ref{fig:solution-quality}. Averaging over the last optimization steps in Figure~\ref{fig:accuracy-most-likely-averaged} shows the stability of the most likely shot at solving the problem.\\
From the results in Figure~\ref{fig:accuracy}, all methods significantly exceed the standard VQE baseline (SVQE), with SHA8 showing a 29.99\% improvement. SHA consistently outperforms Layer-VQE (L-VQE) and layerwise learning (LL) in terms of raw accuracy, given enough partitions. Specifically, SHA8 shows a 17.58\% better mean accuracy than LL and 5.12\% better than L-VQE, demonstrating the effectiveness of our approach. However, the QAOA baseline still performs significantly better than all VQE-based approaches in terms of raw accuracy.\\
When focusing on the most likely shot,the QAOA performs the worst (Figure~\ref{fig:accuracy-most-likely-averaged}), while SHA shows an increasing accuracy trend. These results suggest that the QAOA's state vector contains many superposition states resembling correct solutions but is more spread among incorrect ones. In contrast, VQE-based approaches like L-VQE and LL focus on a pronounced peak, while SHA is more volatile.\\
In conclusion, the best-performing approach depends on the use case and available hardware. If hardware allows,the QAOA is best for overall accuracy in our low layer depth study. Otherwise, training a VQE with SHA yields the best overall accuracy. For a stable peak at a correct solution, layerwise learning-based VQEs perform almost perfectly, while the QAOA performs worse. Future work should investigate scalability for deeper PQCs.

\subsection{Combining SHA with Other Quantum Learning Methods}
\label{subsec:shacombined}
\begin{figure*}[ht]
    \begin{center}
    \begin{subfigure}[t]{0.49\textwidth}
         \centering
         \includegraphics[width=\columnwidth]{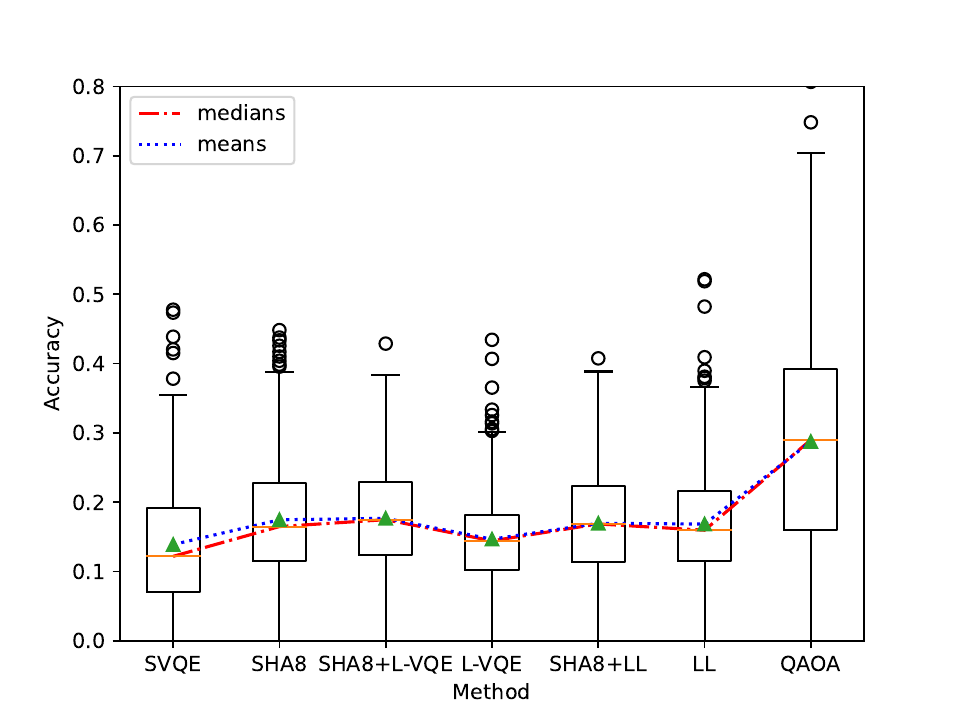}
    \caption{Ratio of correct solutions.}
    \label{fig:accuracy-comb}
     \end{subfigure}
     \begin{subfigure}[t]{0.49\textwidth}
         \centering
         \includegraphics[width=\columnwidth]{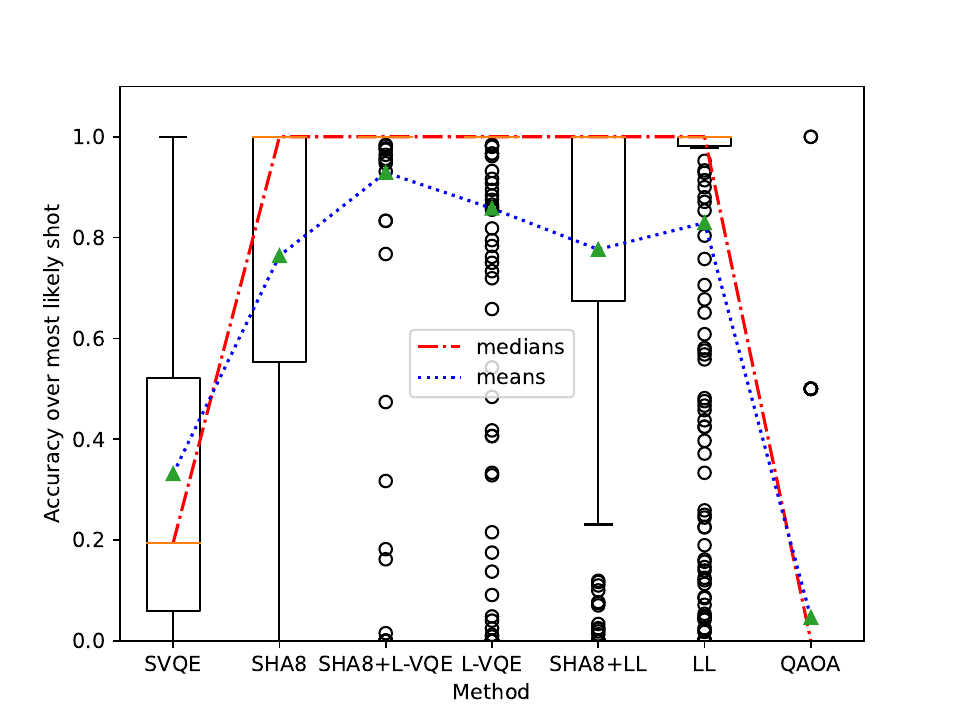}
    \caption{Ratio of correct solutions of the most likely shot, averaged over the last 3\% of optimization steps.}
    \label{fig:accuracy-most-likely-averaged-comb}
     \end{subfigure}
    \caption{Solution quality averaged over all graphs and circuit architectures when combining different approaches. From \cite{Stein_Roshani}}
    \label{fig:solution-quality-comb}
    \end{center}
\end{figure*}

SHA's learning technique, which modifies the cost function, can be combined with layerwise learning approaches that alter the PQC or the set of trainable parameters as well as quantum alghorithms like the QAOA. We evaluate the straightforward approach of using SHA to train each newly added circuit layer, examining overall accuracy and the accuracy of the most likely shot, as shown in Figure~\ref{fig:solution-quality-comb}.\\
Combining SHA8 and L-VQE achieves the best results with a median accuracy improvement of 35.5\% against the standard VQE (SVQE), revealing a powerful synergy. SHA+L-VQE outperforms the previous state-of-the-art VQE approach (LL) by 5\% in the mean. For the most likely shot, the SHA+L-VQE hybrid also performs best, improving 8.31\% over the previous best mean result (L-VQE). Interestingly, SHA+LL performs only slightly better than SHA but worse than LL for this metric, indicating that hybrid approaches do not always enhance performance.\\

\subsection{Combining SHA with the QAOA}

\begin{figure*}[ht]
    \begin{center}
    \begin{subfigure}[t]{0.49\textwidth}
         \centering
         \includegraphics[width=\columnwidth]{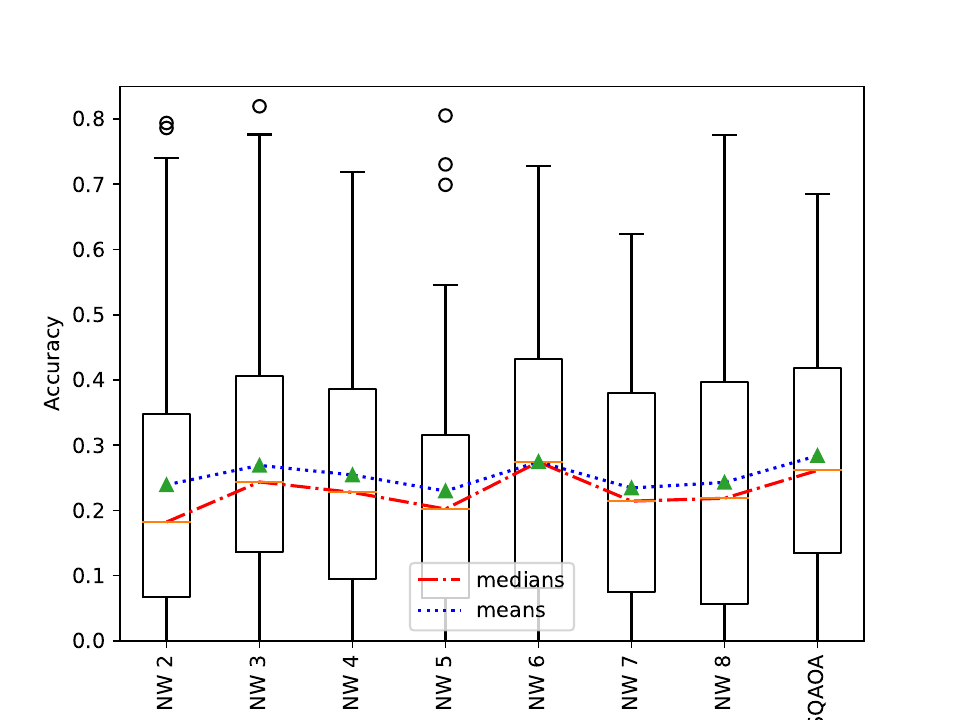}
    \caption{Ratio of correct solutions.}
    \label{fig:QSHA_ac}
    \end{subfigure}
    \begin{subfigure}[t]{0.49\textwidth}
         \centering
         \includegraphics[width=\columnwidth]{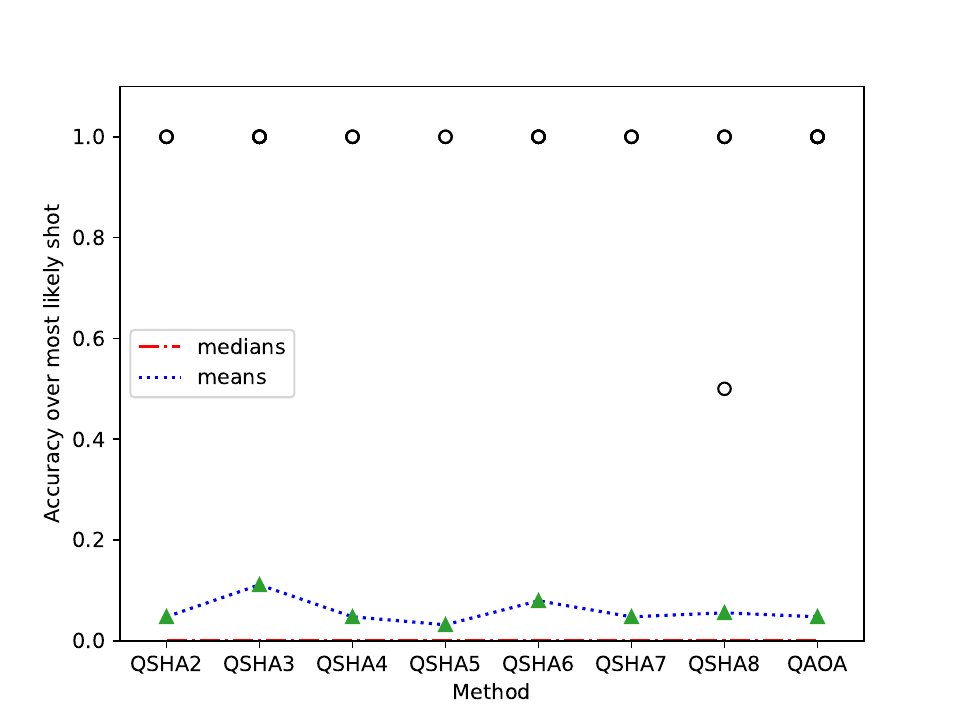}
    \caption{Ratio of correct solutions of the most likely shot, averaged over the last 2\% of optimization steps.}
    \label{fig:QSHA_Acc_most_likley}
     \end{subfigure}
    \caption{Solution quality averaged over all graphs and circuit architectures when combining QAOA and SHA.}
    \label{fig:QAOA_SHA}
    \end{center}
\end{figure*}

As SHA can be combined with other Quantum learning methods as seen in \ref{subsec:shacombined}, it is sensible to combine it with Quantum algorithms like the QAOA. We evaluate the effect of SHA onto the QAOA, which effectively layers both the circuit and cost function into correlated partitions. We examine again the overall accuracy and the accuracy of the most likely shot, as shown in Figure \ref{fig:QAOA_SHA}.\\
For the overall accuracy, we see that the standard QAOA outperforms the QAOA+SHA (NW $i$) in the mean by 3.4\% compared to the best combination of SHA6 with QAOA. However, looking at the median there is an improvement of 4.9\% for the same layer. Interestingly, there is no clear trend as to the other methods, that more SHA layers result into better solution accuracy. Examining the results of the most likely shots shows QAOA+SHA always being equal or better than the standard QAOA with improvements of up to 136\% in the mean. The median on the other hand remains at 0 for all results. Furthermore, compared to the other methods (see Fig. \ref{fig:accuracy-most-likely-averaged} and Fig.\ref{fig:accuracy-most-likely-averaged-comb}) it is still the worst method. Concluding these results, we find that the combination of QAOA and SHA doesn't enhance the results in a meaningful way. Future work will have to investigate if this is also true for larger problems. 

\subsection{Time Complexity}
\begin{figure}[ht]
    \centering
    \includegraphics[width=\columnwidth]{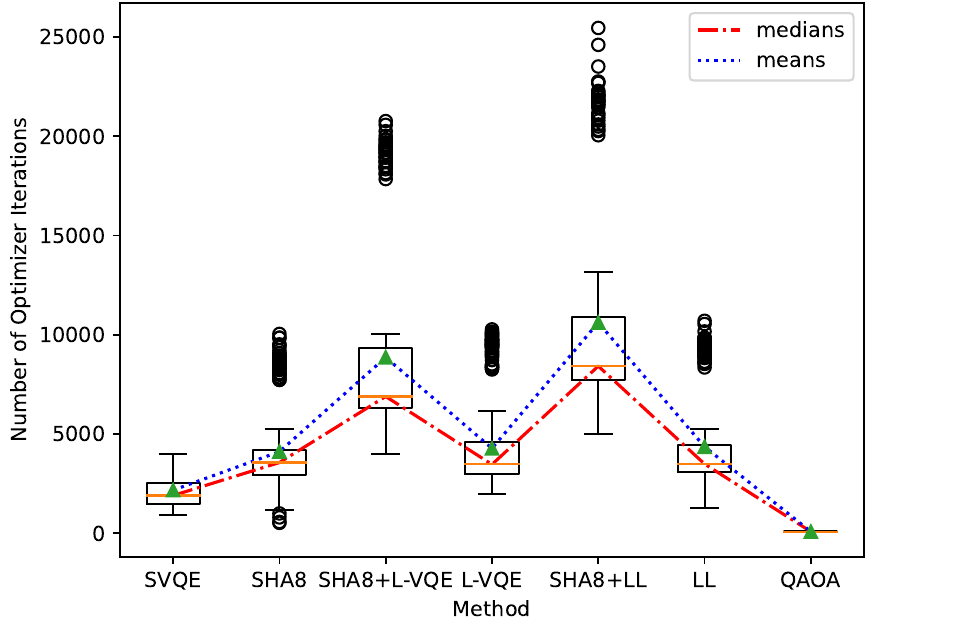}
    \caption{Number of optimization iterations. From \cite{Stein_Roshani}}
    \label{fig:speed-w/pre}
\end{figure}

Training duration is a crucial performance indicator. Fewer optimization steps imply faster training, considering a similar number of parameters. Quantum gradient calculation scales linearly with the number of parameters (Section \ref{subsec:paramtraining}). In layerwise approaches, the number of concurrently trained parameters is smaller, impacting runtime analysis. On average, LL trains about half the parameters in the full PQC, while L-VQE trains roughly $\nicefrac{2}{3}$. Circuit depth also changes in layerwise learning, affecting execution time. Our hyperparameters result in LL executing $\nicefrac{3}{4}$ of the full PQC on average, while L-VQE executes $\nicefrac{2}{3}$.\\
Due to the small number of parameters and layers, we focus on the number of optimization iterations shown in Figure~\ref{fig:speed-w/pre}. Comparing SHA to the standard VQE, we see nearly double the number of optimization iterations, indicating that improved solution quality comes at the cost of longer training. However, in real-world applications, better solution quality may outweigh increased runtime, such as when benchmarking for early quantum advantage. L-VQE and LL will have faster wallclock times than SHA despite similar optimization iterations. For hybrid approaches, a significant runtime increase limits practical use, despite qualitative improvements. SHA+L-VQE, with better solution quality, also trains faster, suggesting an parameter landscape that is easier to navigate. Comparing VQE-based approaches to the QAOA is challenging since the QAOA's PQC has only six parameters, speeding up optimization. This highlights the QAOA's faster trainability, especially with short PQCs.

\subsection{SHA on the Max-Cut problem}

\begin{figure}[ht]
    \centering
    \includegraphics[width=\columnwidth]{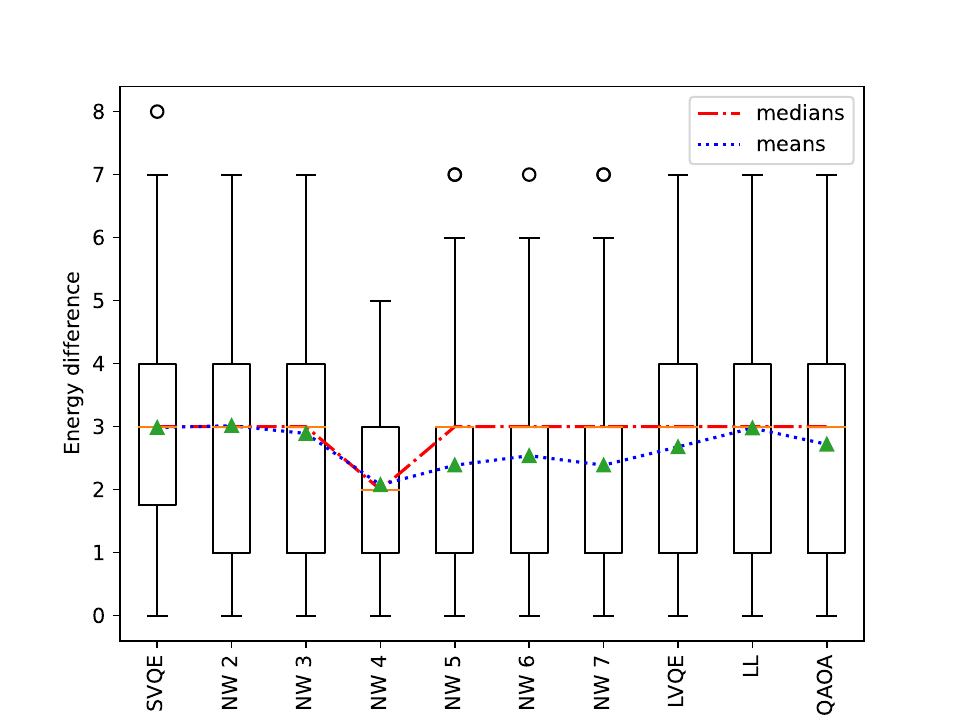}
    \caption{Energy differences averaged over all graphs and circuits for all aproaches}
    \label{fig:maxcut}
\end{figure}

To evaluate the Max-Cut problem, we focus on the differences between the cost of the optimal solution and the result of the corresponding minimization process. As Max-Cut is an optimization problem we are interested in the ability of the respective method to reliably get close to the global minimum instead of the previously used overall accuracy.\\
An examination of the results in Fig. \ref{fig:maxcut} show that SHA is able to outperform the other methods. Against the standard VQE all SHA layers (NW $i$) except NW 2 are exceeding it, with an improvement of up to 43,89 \% for NW 4. This trend continuous also for the methods of Layer-VQE (LVQE), layerwise learning (LL) and the QAOA, although SHA starts to show better results than them starting with NW 4. The boost of the mean compared to the respective methods are 29.08 \% for L-VQE, 43.34 \% for LL and 31,04 \% for the QAOA. The median is due to the comparably low energy of the optimal cut on these particular graphs constant at 3 except for NW 4, which again performs better with a median of 2. Interestingly, LL performs far worse at the Max-Cut problem than it did with the graph coloring problem, barely being better than the standard VQE. To further validate these results we propose further investigations into larger problem instances.

\section{\uppercase{Conclusion}}\label{sec:conclusion}
In this work we extended on our previous case study for the Sequential Hamiltonian Assembly (SHA) technique, which iteratively approximates the potentially global cost function by progressively assembling it from its local components. Previously, we have shown the advantages of SHA over the standard method of the VQE and the state of the art of LL and Layer-VQE, when applied on a graph coloring problem \cite{Stein_Roshani}. In our new experiments we have shown that SHA still performs better than comparative methods when applied onto the Max-Cut problem with improvements of up to 43,89 \% for the VQE and 29,08 \% for the Layer-VQE. This results should be further validated on larger problem instances. Other optimizations problems might also be instructive to investigate with potential candidates being the traveling salesperson or the Knapsack problem. Additionally, we've shown that the choice of the SHA strategy has an impact on the final solution quality. Our newly introduced \emph{cluster} method didn't produce any noteworthy results, however it might be still beneficial to examine the strategy onto larger problem graphs. Furthermore, we have shown that combinations of existing layerwise learning approaches/quantum algorithms with SHA can increase the solution quality even further, but this is highly dependent on the used methods. While the merger of SHA and Layyer-VQE is highly beneficial, QAOA doesn't exhibit any significant improvements. Understanding the reasoning behind this behaviour could lead to a new class of quantum learning methods that are especially targeted towards attacking multiple issues of vanishing gradients concurrently, as, in our case locality and expressiveness. Overall, an extensive hyperparameter search is still in order to find faster and and potentially better results for SHA.

\section*{\uppercase{ACKNOWLEDGEMENTS}}\label{sec:Acknowledgement}

This paper was partially funded by the German Federal Ministry for Economic Affairs and Climate Action through the funding program ``Quantum Computing– Applications for the industry'' based on the allowance ``Development of digital technologies'' (contract number: 01MQ22008A) and is part of the Munich Quantum Valley, which is supported by the Bavarian state government with funds from the Hightech Agenda Bayern Plus.
\clearpage
%
\bibliographystyle{splncs04}
\bibliography{main}

\begin{thebibliography}{10}
\providecommand{\url}[1]{\texttt{#1}}
\providecommand{\urlprefix}{URL }
\providecommand{\doi}[1]{https://doi.org/#1}

\bibitem{10.1109/FOCS.2004.8}
Aharonov, D., van Dam, W., Kempe, J., Landau, Z., Lloyd, S., Regev, O.: Adiabatic quantum computation is equivalent to standard quantum computation. In: Proceedings of the 45th Annual IEEE Symposium on Foundations of Computer Science. p. 42–51. FOCS '04, IEEE Computer Society, USA (2004). \doi{10.1109/FOCS.2004.8}, \url{https://doi.org/10.1109/FOCS.2004.8}

\bibitem{PhysRevX.8.031016}
Albash, T., Lidar, D.A.: Demonstration of a scaling advantage for a quantum annealer over simulated annealing. Phys. Rev. X  \textbf{8},  031016 (Jul 2018). \doi{10.1103/PhysRevX.8.031016}, \url{https://link.aps.org/doi/10.1103/PhysRevX.8.031016}

\bibitem{bansal:LIPIcs:2014:4827}
Bansal, N.: {New Developments in Iterated Rounding (Invited Talk)}. In: Raman, V., Suresh, S.P. (eds.) 34th International Conference on Foundation of Software Technology and Theoretical Computer Science (FSTTCS 2014). Leibniz International Proceedings in Informatics (LIPIcs), vol.~29, pp. 1--10. Schloss Dagstuhl--Leibniz-Zentrum fuer Informatik, Dagstuhl, Germany (2014). \doi{10.4230/LIPIcs.FSTTCS.2014.1}, \url{http://drops.dagstuhl.de/opus/volltexte/2014/4827}

\bibitem{NIPS2006_5da713a6}
Bengio, Y., Lamblin, P., Popovici, D., Larochelle, H.: Greedy layer-wise training of deep networks. In: Sch\"{o}lkopf, B., Platt, J., Hoffman, T. (eds.) Advances in Neural Information Processing Systems. vol.~19. MIT Press (2006), \url{https://proceedings.neurips.cc/paper_files/paper/2006/file/5da713a690c067105aeb2fae32403405-Paper.pdf}

\bibitem{blekos2023review}
Blekos, K., Brand, D., Ceschini, A., Chou, C.H., Li, R.H., Pandya, K., Summer, A.: A review on quantum approximate optimization algorithm and its variants (2023)

\bibitem{Born1928}
Born, M., Fock, V.: {Beweis des Adiabatensatzes}. Zeitschrift f{\"{u}}r Phys.  \textbf{51}(3),  165--180 (1928). \doi{10.1007/BF01343193}, \url{https://doi.org/10.1007/BF01343193}

\bibitem{bradley2010learning}
Bradley, D.M.: Learning in modular systems. Carnegie Mellon University (2010)

\bibitem{PhysRevA.103.032607}
Campos, E., Nasrallah, A., Biamonte, J.: Abrupt transitions in variational quantum circuit training. Phys. Rev. A  \textbf{103},  032607 (Mar 2021). \doi{10.1103/PhysRevA.103.032607}, \url{https://link.aps.org/doi/10.1103/PhysRevA.103.032607}

\bibitem{Cerezo2021}
Cerezo, M., Arrasmith, A., Babbush, R., Benjamin, S.C., Endo, S., Fujii, K., McClean, J.R., Mitarai, K., Yuan, X., Cincio, L., Coles, P.J.: {Variational quantum algorithms}. Nat. Rev. Phys.  \textbf{3}(9),  625--644 (2021). \doi{10.1038/s42254-021-00348-9}, \url{https://doi.org/10.1038/s42254-021-00348-9}

\bibitem{cerezo2021cost}
Cerezo, M., Sone, A., Volkoff, T., Cincio, L., Coles, P.J.: Cost function dependent barren plateaus in shallow parametrized quantum circuits. Nature communications  \textbf{12}(1), ~1791 (2021)

\bibitem{deutsch1992rapid}
Deutsch, D., Jozsa, R.: Rapid solution of problems by quantum computation. Proceedings of the Royal Society of London. Series A: Mathematical and Physical Sciences  \textbf{439}(1907),  553--558 (1992)

\bibitem{Du2022}
Du, Y., Huang, T., You, S., Hsieh, M.H., Tao, D.: {Quantum circuit architecture search for variational quantum algorithms}. npj Quantum Inf.  \textbf{8}(1), ~62 (2022). \doi{10.1038/s41534-022-00570-y}, \url{https://doi.org/10.1038/s41534-022-00570-y}

\bibitem{farhi2014quantum}
Farhi, E., Goldstone, J., Gutmann, S.: A quantum approximate optimization algorithm (2014)

\bibitem{fontana2023adjoint}
Fontana, E., Herman, D., Chakrabarti, S., Kumar, N., Yalovetzky, R., Heredge, J., Sureshbabu, S.H., Pistoia, M.: The adjoint is all you need: Characterizing barren plateaus in quantum ans\"atze (2023)

\bibitem{10.1214/aoms/1177706098}
Gilbert, E.N.: {Random Graphs}. The Annals of Mathematical Statistics  \textbf{30}(4),  1141 -- 1144 (1959). \doi{10.1214/aoms/1177706098}, \url{https://doi.org/10.1214/aoms/1177706098}

\bibitem{pmlr-v9-glorot10a}
Glorot, X., Bengio, Y.: Understanding the difficulty of training deep feedforward neural networks. In: Teh, Y.W., Titterington, M. (eds.) Proceedings of the Thirteenth International Conference on Artificial Intelligence and Statistics. Proceedings of Machine Learning Research, vol.~9, pp. 249--256. PMLR, Chia Laguna Resort, Sardinia, Italy (13--15 May 2010), \url{https://proceedings.mlr.press/v9/glorot10a.html}

\bibitem{10.1145/237814.237866}
Grover, L.K.: A fast quantum mechanical algorithm for database search. In: Proceedings of the Twenty-Eighth Annual ACM Symposium on Theory of Computing. p. 212–219. STOC '96, Association for Computing Machinery, New York, NY, USA (1996). \doi{10.1145/237814.237866}, \url{https://doi.org/10.1145/237814.237866}

\bibitem{hagberg2008exploring}
Hagberg, A., Swart, P., S~Chult, D.: Exploring network structure, dynamics, and function using networkx. Tech. rep., Los Alamos National Lab.(LANL), Los Alamos, NM (United States) (2008)

\bibitem{PRXQuantum.3.010313}
Holmes, Z., Sharma, K., Cerezo, M., Coles, P.J.: Connecting ansatz expressibility to gradient magnitudes and barren plateaus. PRX Quantum  \textbf{3},  010313 (Jan 2022). \doi{10.1103/PRXQuantum.3.010313}, \url{https://link.aps.org/doi/10.1103/PRXQuantum.3.010313}

\bibitem{9345015}
Huang, Y., Lei, H., Li, X.: An empirical study of optimizers for quantum machine learning. In: 2020 IEEE 6th International Conference on Computer and Communications (ICCC). pp. 1560--1566 (2020). \doi{10.1109/ICCC51575.2020.9345015}

\bibitem{IKOTUN2023178}
Ikotun, A.M., Ezugwu, A.E., Abualigah, L., Abuhaija, B., Heming, J.: K-means clustering algorithms: A comprehensive review, variants analysis, and advances in the era of big data. Information Sciences  \textbf{622},  178--210 (2023). \doi{https://doi.org/10.1016/j.ins.2022.11.139}, \url{https://www.sciencedirect.com/science/article/pii/S0020025522014633}

\bibitem{Joshi_2021}
Joshi, N., Katyayan, P., Ahmed, S.A.: Evaluating the performance of some local optimizers for variational quantum classifiers. Journal of Physics: Conference Series  \textbf{1817}(1),  012015 (mar 2021). \doi{10.1088/1742-6596/1817/1/012015}, \url{https://dx.doi.org/10.1088/1742-6596/1817/1/012015}

\bibitem{kashif2023impact}
Kashif, M., Al-Kuwari, S.: The impact of cost function globality and locality in hybrid quantum neural networks on nisq devices. Machine Learning: Science and Technology  \textbf{4}(1),  015004 (2023)

\bibitem{knill2007optimal}
Knill, E., Ortiz, G., Somma, R.D.: Optimal quantum measurements of expectation values of observables. Physical Review A  \textbf{75}(1),  012328 (2007)

\bibitem{lanczos2012variational}
Lanczos, C.: The Variational Principles of Mechanics. Dover Books on Physics, Dover Publications (2012), \url{https://books.google.de/books?id=cmPDAgAAQBAJ}

\bibitem{9669165}
Liu, X., Angone, A., Shaydulin, R., Safro, I., Alexeev, Y., Cincio, L.: Layer vqe: A variational approach for combinatorial optimization on noisy quantum computers. IEEE Transactions on Quantum Engineering  \textbf{3},  1--20 (2022). \doi{10.1109/TQE.2021.3140190}

\bibitem{10.3389/fphy.2014.00005}
Lucas, A.: Ising formulations of many np problems. Frontiers in Physics  \textbf{2} (2014). \doi{10.3389/fphy.2014.00005}, \url{https://www.frontiersin.org/articles/10.3389/fphy.2014.00005}

\bibitem{McClean2018}
McClean, J.R., Boixo, S., Smelyanskiy, V.N., Babbush, R., Neven, H.: {Barren plateaus in quantum neural network training landscapes}. Nat. Commun.  \textbf{9}(1), ~4812 (2018). \doi{10.1038/s41467-018-07090-4}, \url{https://doi.org/10.1038/s41467-018-07090-4}

\bibitem{PhysRevA.98.032309}
Mitarai, K., Negoro, M., Kitagawa, M., Fujii, K.: Quantum circuit learning. Phys. Rev. A  \textbf{98},  032309 (Sep 2018). \doi{10.1103/PhysRevA.98.032309}, \url{https://link.aps.org/doi/10.1103/PhysRevA.98.032309}

\bibitem{nielsen_chuang_2010}
Nielsen, M.A., Chuang, I.L.: Quantum Computation and Quantum Information: 10th Anniversary Edition. Cambridge University Press (2010). \doi{10.1017/CBO9780511976667}

\bibitem{scikit-learn}
Pedregosa, F., Varoquaux, G., Gramfort, A., Michel, V., Thirion, B., Grisel, O., Blondel, M., Prettenhofer, P., Weiss, R., Dubourg, V., Vanderplas, J., Passos, A., Cournapeau, D., Brucher, M., Perrot, M., Duchesnay, E.: Scikit-learn: Machine learning in {P}ython. Journal of Machine Learning Research  \textbf{12},  2825--2830 (2011)

\bibitem{Peruzzo2014}
Peruzzo, A., McClean, J., Shadbolt, P., Yung, M.H., Zhou, X.Q., Love, P.J., Aspuru-Guzik, A., O'Brien, J.L.: {A variational eigenvalue solver on a photonic quantum processor}. Nat. Commun.  \textbf{5}(1), ~4213 (2014). \doi{10.1038/ncomms5213}, \url{https://doi.org/10.1038/ncomms5213}

\bibitem{pirnay2023inprinciple}
Pirnay, N., Ulitzsch, V., Wilde, F., Eisert, J., Seifert, J.P.: An in-principle super-polynomial quantum advantage for approximating combinatorial optimization problems (2023)

\bibitem{COBYLA1}
Powell, M.J.D.: A Direct Search Optimization Method That Models the Objective and Constraint Functions by Linear Interpolation, pp. 51--67. Springer Netherlands, Dordrecht (1994). \doi{10.1007/978-94-015-8330-5}, \url{https://doi.org/10.1007/978-94-015-8330-5}

\bibitem{ragone2023unified}
Ragone, M., Bakalov, B.N., Sauvage, F., Kemper, A.F., Marrero, C.O., Larocca, M., Cerezo, M.: A unified theory of barren plateaus for deep parametrized quantum circuits (2023)

\bibitem{PRXQuantum.3.020365}
Sack, S.H., Medina, R.A., Michailidis, A.A., Kueng, R., Serbyn, M.: Avoiding barren plateaus using classical shadows. PRX Quantum  \textbf{3},  020365 (Jun 2022). \doi{10.1103/PRXQuantum.3.020365}, \url{https://link.aps.org/doi/10.1103/PRXQuantum.3.020365}

\bibitem{PhysRevA.99.032331}
Schuld, M., Bergholm, V., Gogolin, C., Izaac, J., Killoran, N.: Evaluating analytic gradients on quantum hardware. Phys. Rev. A  \textbf{99},  032331 (Mar 2019). \doi{10.1103/PhysRevA.99.032331}, \url{https://link.aps.org/doi/10.1103/PhysRevA.99.032331}

\bibitem{PhysRevA.103.032430}
Schuld, M., Sweke, R., Meyer, J.J.: Effect of data encoding on the expressive power of variational quantum-machine-learning models. Phys. Rev. A  \textbf{103},  032430 (Mar 2021). \doi{10.1103/PhysRevA.103.032430}, \url{https://link.aps.org/doi/10.1103/PhysRevA.103.032430}

\bibitem{sim2019expressibility}
Sim, S., Johnson, P.D., Aspuru-Guzik, A.: Expressibility and entangling capability of parameterized quantum circuits for hybrid quantum-classical algorithms. Advanced Quantum Technologies  \textbf{2}(12),  1900070 (2019)

\bibitem{https://doi.org/10.1002/qute.201900070}
Sim, S., Johnson, P.D., Aspuru-Guzik, A.: Expressibility and entangling capability of parameterized quantum circuits for hybrid quantum-classical algorithms. Advanced Quantum Technologies  \textbf{2}(12),  1900070 (2019). \doi{https://doi.org/10.1002/qute.201900070}, \url{https://onlinelibrary.wiley.com/doi/abs/10.1002/qute.201900070}

\bibitem{Skolik2021}
Skolik, A., McClean, J.R., Mohseni, M., van~der Smagt, P., Leib, M.: {Layerwise learning for quantum neural networks}. Quantum Mach. Intell.  \textbf{3}(1), ~5 (2021). \doi{10.1007/s42484-020-00036-4}, \url{https://doi.org/10.1007/s42484-020-00036-4}

\bibitem{Stein_Roshani}
Stein, J., Roshani, N., Zorn, M., Altmann, P., Kölle, M., Linnhoff-Popien, C.: Improving parameter training for vqes by sequential hamiltonian assembly. In: Proceedings of the 16th International Conference on Agents and Artificial Intelligence. SCITEPRESS - Science and Technology Publications (2024). \doi{10.5220/0012312500003636}, \url{http://dx.doi.org/10.5220/0012312500003636}

\bibitem{stilck2021limitations}
Stilck~Fran{\c{c}}a, D., Garcia-Patron, R.: Limitations of optimization algorithms on noisy quantum devices. Nature Physics  \textbf{17}(11),  1221--1227 (2021)

\bibitem{9259934}
Tabi, Z., El-Safty, K.H., Kallus, Z., Hága, P., Kozsik, T., Glos, A., Zimborás, Z.: Quantum optimization for the graph coloring problem with space-efficient embedding. In: 2020 IEEE International Conference on Quantum Computing and Engineering (QCE). pp. 56--62 (2020). \doi{10.1109/QCE49297.2020.00018}

\bibitem{uvarov2021barren}
Uvarov, A., Biamonte, J.D.: On barren plateaus and cost function locality in variational quantum algorithms. Journal of Physics A: Mathematical and Theoretical  \textbf{54}(24),  245301 (2021)

\bibitem{wang2021noise}
Wang, S., Fontana, E., Cerezo, M., Sharma, K., Sone, A., Cincio, L., Coles, P.J.: Noise-induced barren plateaus in variational quantum algorithms. Nature communications  \textbf{12}(1), ~6961 (2021)

\bibitem{PhysRevE.76.031131}
Zdeborov\'a, L., Krzk{a}ka\l{}a, F.: Phase transitions in the coloring of random graphs. Phys. Rev. E  \textbf{76},  031131 (Sep 2007). \doi{10.1103/PhysRevE.76.031131}, \url{https://link.aps.org/doi/10.1103/PhysRevE.76.031131}

\bibitem{NEURIPS2022_7611a3cb}
Zhang, K., Liu, L., Hsieh, M.H., Tao, D.: Escaping from the barren plateau via gaussian initializations in deep variational quantum circuits. In: Koyejo, S., Mohamed, S., Agarwal, A., Belgrave, D., Cho, K., Oh, A. (eds.) Advances in Neural Information Processing Systems. vol.~35, pp. 18612--18627. Curran Associates, Inc. (2022), \url{https://proceedings.neurips.cc/paper_files/paper/2022/file/7611a3cb5d733e628081431445cb01fd-Paper-Conference.pdf}

\end{thebibliography}

\end{document}